\documentclass[conference]{IEEEtran}
\makeatletter
\def\ps@headings{%
\def\@oddhead{\mbox{}\scriptsize\rightmark \hfil \thepage}%
\def\@evenhead{\scriptsize\thepage \hfil \leftmark\mbox{}}%
\def\@oddfoot{}%
\def\@evenfoot{}}
\makeatother
\usepackage[cmex10]{amsmath}
\usepackage{array}
\usepackage{url}
\usepackage{makecell}
\usepackage{graphicx}
\graphicspath{{./figures/}}

\begin{document}
%
\title{Phoenix Cloud: Consolidating Different Computing Loads on Shared Cluster System for Large Organization}



%
\author{\IEEEauthorblockN{Jianfeng Zhan\IEEEauthorrefmark{1,2},
Lei Wang\IEEEauthorrefmark{1},
Bibo Tu\IEEEauthorrefmark{1},
Yong Li\IEEEauthorrefmark{1},
Peng Wang\IEEEauthorrefmark{1},
Wei Zhou\IEEEauthorrefmark{1} and
Dan Meng\IEEEauthorrefmark{1}}
\IEEEauthorblockA{\IEEEauthorrefmark{1}Institute of Computing Technology,Chinese Academy of Sciences Beijing, China 100190}
\IEEEauthorblockA{\IEEEauthorrefmark{2}
Email: jfzhan@ncic.ac.cn}
}


\maketitle

\begin{abstract}

Different departments of a large organization often run dedicated cluster systems for different computing loads, like HPC (high performance computing) jobs or Web service applications. In this paper, we have designed and implemented a cloud management system software \emph{Phoenix Cloud} to consolidate heterogeneous workloads from different departments affiliated to the same organization on the shared cluster system. We have also proposed cooperative resource provisioning and management policies for a large organization and its affiliated departments, running HPC jobs and Web service applications, to share the consolidated cluster system. The experiments show that in comparison with the case that each department operates its dedicated cluster system, \emph{Phoenix Cloud} significantly decreases the scale of the required cluster system for a large organization, improves the benefit of the scientific computing department, and at the same time provisions enough resources to the other department running Web services with varying loads.
\footnote{This document is dated from August 13, 2008 and contains an early summary of experiences of this project, which also available from the web site of \emph{the First Workshop of Cloud Computing and its Application (http://www.cca08.org/papers/Poster-8-Jianfeng-Zhan.pdf)}. The extended version with the title of \emph{PhoenixCloud: Provisioning Resources for Heterogeneous Workloads in Cloud Computing} can be downloaded from http://arxiv.org/abs/1006.1401.}
\end{abstract}

\section{Introduction}
Since 2007, a client from a large organization, which we keep anonymous at its request, requires us to build the system software for managing a shared infrastructure. This large organization has two representative departments: \emph{one running a batch queuing system for HPC jobs}, and \emph{the other one responsible of providing Web services}, of which the ratios of peak loads to normal loads are high. So two representative departments from this big organization have operated two cluster systems with independent administration staffs and found many annoying problems: first, resource utilization rates of two cluster systems are varying. For peak loads, dedicated cluster systems can not provision enough resources, while for normal loads lots of resources are idle; second, the number of administration staffs for two separated cluster systems is high. The client inquired us whether it is possible to help them consolidate two cluster systems on one shared system.

At same time, we have noticed that many famous IT companies are advocating and experiencing cloud computing. For example, Amazon \cite{1} has provided cloud computing services like \emph{elastic computing cloud (EC2)} and
\emph{simple storage service (S3)} to end users. What is the link between services provided by Amazon and the
requirement of our anonymous client? In our opinion, driven by the cost, cloud computing is a new wave of
reconstructing and consolidating data centers. Traditional cluster system software is self-containing \cite{2}, inadequate for adapting to this change. EC2 and S3 are big efforts to provide virtualized hosting environments for end users, but it can not provide the customized system stack software to consolidate heterogeneous loads on shared cluster systems for large organizations. In fact, there lies no one-fit-all solution.

In this paper, we focus on developing cloud computing management software that enables the consolidation of
heterogeneous workloads on shared cluster systems for large organizations, and we stress that \emph{we do not target the design of capability-oriented system software stack} \cite{3}. To the best of our knowledge, this is the first paper to propose the layered architecture of cloud computing management software for large organizations that intend to consolidate heterogeneous workloads from different departments on shared cluster systems. \cite{4} proposes the utility computing service framework to facilitate code reuse in the context of traditional data centers, but do not consider how to enable consolidating different types of workloads. \cite{5} \cite{6} propose \emph{Cluster on Demand}(COD) as a new mechanism for dynamical cluster resource management in the contexts of Internet hosting center \cite{5} or scientific computing \cite{6}, but their works mainly focus on dynamic resource provisioning in respective computing contexts.

The distinguished differences of our system and architecture from others are that: first, we develop a common service framework as a foundation for cloud computing system software; second, with the support of a common service framework, we create cloud management services respectively for scientific computing (HPC jobs) and web
service applications; third, we propose optimal resource management and provision policies for heterogeneous workloads to cooperatively share cluster resources. The contribution of this paper can be concluded as:
\begin{itemize}
  \item We have designed and implemented a cloud management system software \emph{Phoenix Cloud} with the \emph{layered architecture} to consolidated HPC jobs and Web service applications on shared cluster systems.
  \item We have proposed cooperative resource provisioning and management policies for a large organization and its affiliated departments to share the cluster system.
  \item Our experiments show that in comparison with the case that each department maintains its dedicated cluster system, consolidating HPC jobs and Web service applications with cooperative resource provisioning and management policies can significantly decrease the scale of the required cluster system for a large organization, at the same time improve the benefit of the scientific computing department while provisioning enough resources to the department who runs Web service applications with varying loads.
\end{itemize}

Our paper includes four sections. In Section \ref{design_implementation}, we explain the design and implementation issue of \emph{Phoenix Cloud}. In Section \ref{evaluation}, we evaluate our system. In Section \ref{conclusion}, we draw a conclusion.

\section{Design and Implementation Issues} \label{design_implementation}
In Section \ref{layered_architecture}, we introduce the layered architecture of \emph{Phoenix Cloud}. In Section \ref{cooperation}, we propose cooperative resource provision and management policies for \emph{Phoenix Cloud}.

\subsection{The Layered Architecture of Phoenix Cloud} \label{layered_architecture}
We divide our cloud computing management software into three independent layers: \emph{shared infrastructure for the resource provider}, \emph{cloud management services for service providers who are different departments affiliated to the same large organization}, and \emph{client tools for end user}. Fig. \ref{architecture} shows the macro-level architecture of our innovative system as follows:

\begin{figure}[h]
\centering
\includegraphics[width=3.5in]{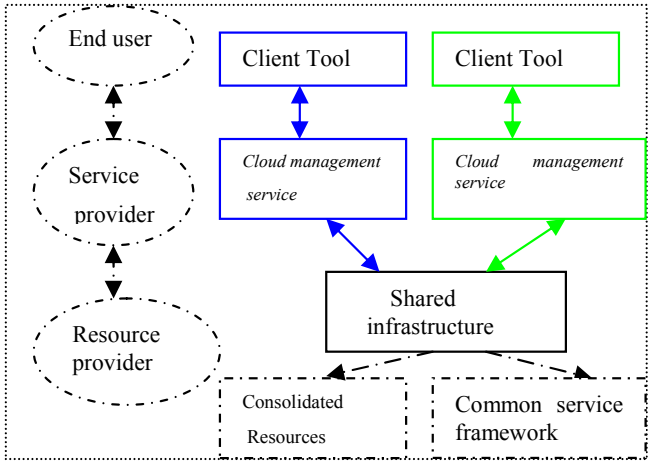}
\caption{The layered architecture of \emph{Phoenix Cloud}.}
\label{architecture}
\end{figure}

\begin{itemize}
  \item The resource provider is responsible for operating \emph{the shared infrastructure}, including \emph{the shared cluster resources} and \emph{the common service framework}. The shared cluster resources include hardware resources, e.g. CPU, memory, and system software like host operating systems. \emph{The common service framework} provides a set of services that manage, monitor the shared cluster resources and provision resources to \emph{cloud management services} for different service providers.
  \item \emph{The cloud management service} (CMS) is a management service for a specific computing load, the implementation detail of which is seen in Fig. \ref{implementation_CMS}.
  \item Client tools: end users use client tools to access services or submit jobs.
\end{itemize}

\begin{figure}[h]
\centering
\includegraphics[width=3.5in]{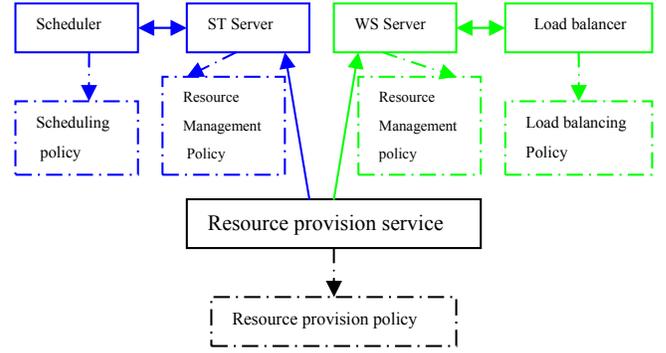}
\caption{The micro-level architecture of \emph{Phoenix Cloud}.}
\label{micro_level}
\end{figure}

Fig. \ref{micro_level} shows the micro-level architecture of \emph{Phoenix Cloud} when two \emph{cloud management services} share a cluster system and reuse \emph{the common service framework}.

One is \emph{a cloud management service for scientific computing like HPC jobs)} (in short, \emph{ST CMS}), including \emph{ST Server} and \emph{Scheduler}, and the other is \emph{a cloud management service for Web services} (\emph{in short, WS CMS}), including \emph{WS Server} and \emph{Load balancer}:
\begin{itemize}
  \item Among \emph{the common service framework}, a service named \emph{Resource Provision Service} with \emph{the customized resource provisioning policy} acts as the proxy of a large organization, responsible for managing and provisioning resource to \emph{different cloud management services}.
  \item \emph{The resource provision policy} determines when \emph{Resource Provision Service} will provision how many resources to different \emph{cloud management services} in what priority.
  \item The \emph{cloud management service} with \emph{a customized resource management policy} and \emph{a scheduling/load balancing policy} behaves as the representative of a service provider, responsible for managing resources, scheduling jobs or distributing requests for load balancing.
  \item \emph{The resource management policy} of a service provider determines when \emph{ST Server} or \emph{WS Server}obtains or returns how many resources to \emph{Resource Provision Service} according to what criteria.
  \item \emph{The scheduling policy} determines \emph{Scheduler} of \emph{ST CMS} when and how to choose HPC jobs for running. \emph{The load balancing policy} determines \emph{Load Balancer} of \emph{WS CMS} how to distribute requests and adjust Web service instances according to what criteria.
\end{itemize}

\begin{figure}[h]
\centering
\includegraphics[width=3.5in]{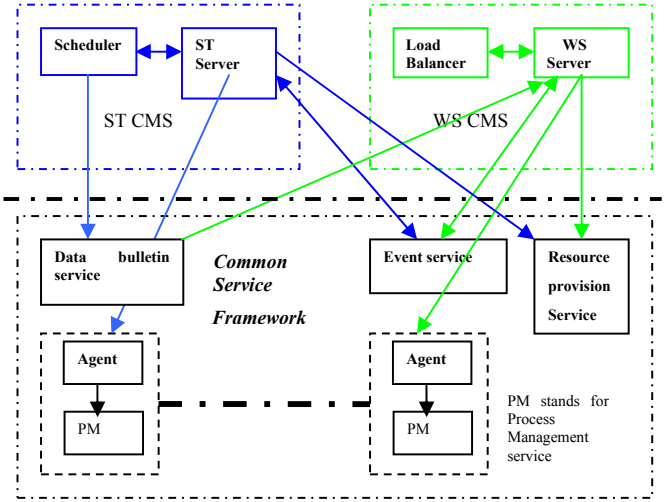}
\caption{With \emph{Phoenix Cloud}, two cloud management services reuse and share the common service framework.}
\label{two_CMS}
\end{figure}

Our innovative system evolves from our previous work \emph{Phoenix}, which is a cluster system software stack \cite{8} \cite{9}. Based on \emph{Phoenix}, we have consolidated two different \emph{cloud management services} respectively for HPC jobs and Web services on the shared cluster system. Fig.\ref{two_CMS} shows the architecture of \emph{ST CMS} and \emph{WS CMS} based on \emph{Phoenix}. The function of \emph{ST CMS} is similar to OpenPBS \cite{10}, while the function of \emph{WM CMS} is similar to the Oceano \cite{11}. But distinguished differences of our systems are as follows:
\begin{itemize}
  \item Different heterogeneous workloads can be consolidated on the shared system;
  \item Different cloud management services can reuse the common service framework;
  \item Different cloud management services can cooperatively share resources under varying loads according to the cooperative policies proposed in section \ref{cooperation}.
\end{itemize}

\subsection{Cooperative Resource Provision and Management Policies} \label{cooperation}
In this section, we propose cooperative resource provisioning and management policies for a large organization
and its affiliated departments. As shown in Fig.\ref{micro_level}, we could specify \emph{the resource provisioning policy} for \emph{Resource Provision Service}, and different resource management policies for \emph{ST Server} and \emph{WS Server}. \emph{The resource provisioning policy} is as follows:
\begin{itemize}
  \item The resource demands from \emph{WS Server} have higher priority than that of \emph{ST Server}.
  \item If there are idle resources for \emph{Resource Provision Service}, it will provision all idle resources to \emph{ST Server}.
  \item If \emph{WS Server} claims urgent resources, \emph{Resource Provision Service} will force \emph{ST Server} to return resources with the size claimed by \emph{WS Server} and then reallocate to \emph{WS Server}.
\end{itemize}

\emph{The resource management policy} of \emph{ST server} is as follows:
\begin{itemize}
  \item \emph{ST Server} passively receives resources provisioned by \emph{Resource Provision Service}.
  \item If \emph{Resource Provision Service} forces \emph{ST Server} to return resources, the latter will release resources immediately with the size demanded by the former.
  \item If there are no enough idle resources for \emph{ST Server}, it will kill jobs in turn from the beginning of job with minimum size and shortest running time, and release enough resources to \emph{Resource Provision Service}.
\end{itemize}

\emph{The resource management policy} of \emph{WS Server} is as follows:
\begin{itemize}
  \item If \emph{WS Server} owns idle resources, it will release them to \emph{Resource Provision Service} immediately. If \emph{WS Server} needs more resources, it will request enough resources from \emph{Resource Provision Service}.
\end{itemize}

\section{EVALUATION AND DISCUSSION} \label{evaluation}

In this section, we will demonstrate that with the cooperative resource provision and management policies
proposed in Section \ref{cooperation}, consolidating HPC jobs and Web services on the shared cluster system, of which we call \emph{dynamic configuration}, can decrease the cost of a large organization in term of \emph{resource consumption} in comparison with \emph{the static configuration}, of which each department maintains its dedicated cluster system.

\subsection{The benefit and cost models}
For a large organization, we use \emph{the size of nodes} to measure the cost of owning a cluster system.
For HPC jobs, we use \emph{the number of completed jobs} to measure the benefit of a service provider; at the same time, we use \emph{the reciprocal of the average turnaround time per job} to measure the benefit of end user.
For Web service application, we use \emph{the throughput} in term of \emph{request/second} to measure the benefit of a service provider; at the same time we use \emph{the average response time of requests} to measure the benefit of an end user.

\subsection{Experiment method and load traces} \label{trace}
Our experiments include two parts: first, in Section \ref{resource_consumption} we obtain the real resource consumption of a Web service application under varying loads on the testbed. Second, based on the real resource consumption of a Web service application obtained in Section \ref{resource_consumption}, we use a simulation method to obtain the resource consumption in the case of consolidating different computing loads from different departments of a large organization on the shared cluster system.
The synthetic request trace of Web service application is obtained from the real trace of World Cup of two
week from June 7 in 1998 \cite{12} with a scaling factor of 2.22. For the World Cup trace, the ratio of the peak load to the normal load is high.
HPC trace is the real trace of SDSC BLUE of two weeks from Apr 25 15:00:03 PDT 2000 on the web
site of http://www.cs.huji.ac.il/labs/parallel/workload/logs.html.

\subsection{The resource consumption of Web service under varying load} \label{resource_consumption}
The testbed is as follows: All nodes are connected with a 1 Gb/s switch. Each node has the same configuration: 8 $\times$ Intel(R) Xeon(R) (2.00GHz) CPU and 2G memory with 64 bit Linux with kernel of 2.6.18-xen. On each node, we deploy eight XEN \cite{13} virtual machines. The configuration of XEN virtual machine is: 1 $\times$ Intel(R) Xeon(R)(2.00GHz) CPU; 256M memory; the guest operating system is 64 bit CentOS with kernel version of 2.6.18.

\begin{figure}[h]
\centering
\includegraphics[width=3.0in]{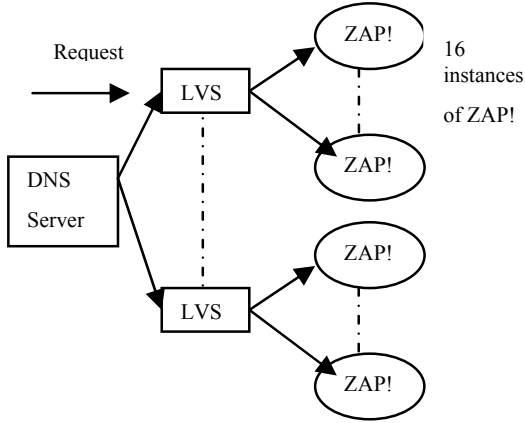}
\caption{The system deployment diagram.}
\label{deployment}
\end{figure}

Fig. \ref{deployment} shows the system deployment diagram. We choose httperf \cite{14} as the load generator, LVS\cite{15} with direct route mode is responsible for distributing requests to the Web service with the least-connection scheduling policy. The DNS server is responsible for distributing connection from each user to one of four LVS with round robin policy. We choose open source software ZAP! \cite{7} as a typical Web service, and each instance of ZAP! is deployed on a virtual machine.

\begin{figure}[h]
\centering
\includegraphics[width=3.5in]{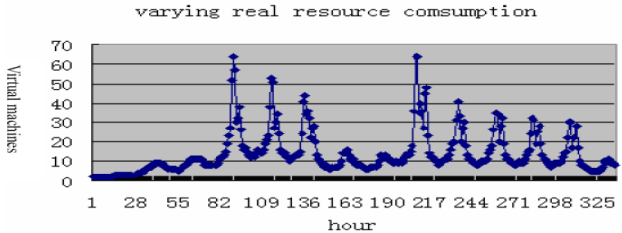}
\caption{The resource consumption of Web service trace in two weeks.}
\label{varying_resource}
\end{figure}

\emph{WS Server} adjusts \emph{the number of instances of Web services} according to the criterion in terms of  \emph{the average utilization rate of CPU consumed by Web service instances}. We presume the number of current instances of information service is $n$. If \emph{the average utilization rate of CPUs consumed by Web service instances} exceeds 80\% in the past 20 seconds, \emph{WS Server} will increase one instance. If \emph{the average utilization rate of CPUs consumed by Web service instances} is lower than $80\% (n-1)/n$ in the past 20 seconds, \emph{WS Server} will decrease one instance until the number of the current instances is equal to 1.
We use the Web service trace described in Section \ref{trace}, and Fig.\ref{varying_resource} shows the varying  resource consumption in two weeks, of which the peak resource demand is 64 virtual machines.

\subsection{The simulation experiments of consolidating computing loads}
\begin{figure}[h]
\centering
\includegraphics[width=3.5in]{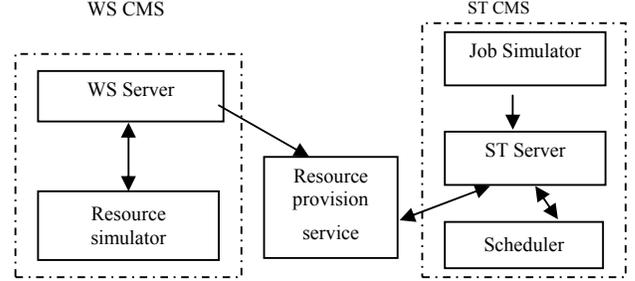}
\caption{The architecture of the simulated system.}
\label{simulated_system}
\end{figure}

We use a simulation method to verify the advantage of consolidating different computing loads from different
departments of a large organization on the shared cluster system. Fig. \ref{simulated_system} shows the architecture of our simulation system, which includes \emph{one cloud management service for scientific computing} (\emph{ST CMS}) and \emph{one cloud management service for Web service} (\emph{WS CMS}). In comparison with the real \emph{Phoenix Cloud} system, our simulated system maintains \emph{Resource Provision Service}, \emph{WS Server}, \emph{ST Server} and \emph{Scheduler}, while other services are removed or substituted.

For \emph{WS CMS}, a daemon named \emph{Resource Simulator} will simulate the varying resource demand of \emph{WS CMS} and drive \emph{WS Server} to obtain or release resources from and to \emph{Resource Provision Service}. We use the real resource consumption in Fig. \ref{resource_consumption} as the input to \emph{Resource Simulator}.
For \emph{ST CMS}, \emph{Scheduler} is specified with \emph{the First-Fit} scheduling policy, and \emph{Job Simulator} is used to simulate the process of submitting jobs. To accelerate the experiment, we speed up the submission and completion of jobs by a factor of 100. This speedup allows two weeks trace to complete in about three hours. The HPC trace is introduced in Section \ref{trace}.
We presume that the software package of Web service are pre-deployed on those reallocated nodes, so the
time of reallocating nodes from \emph{ST Server} to \emph{WS server} is only seconds, includes \emph{the time of killing jobs} and \emph{the time of communicating among \emph{WS Server}, \emph{ST Server} and \emph{Resource Provision Service}}.

\begin{figure}[h]
\centering
\includegraphics[width=3.5in]{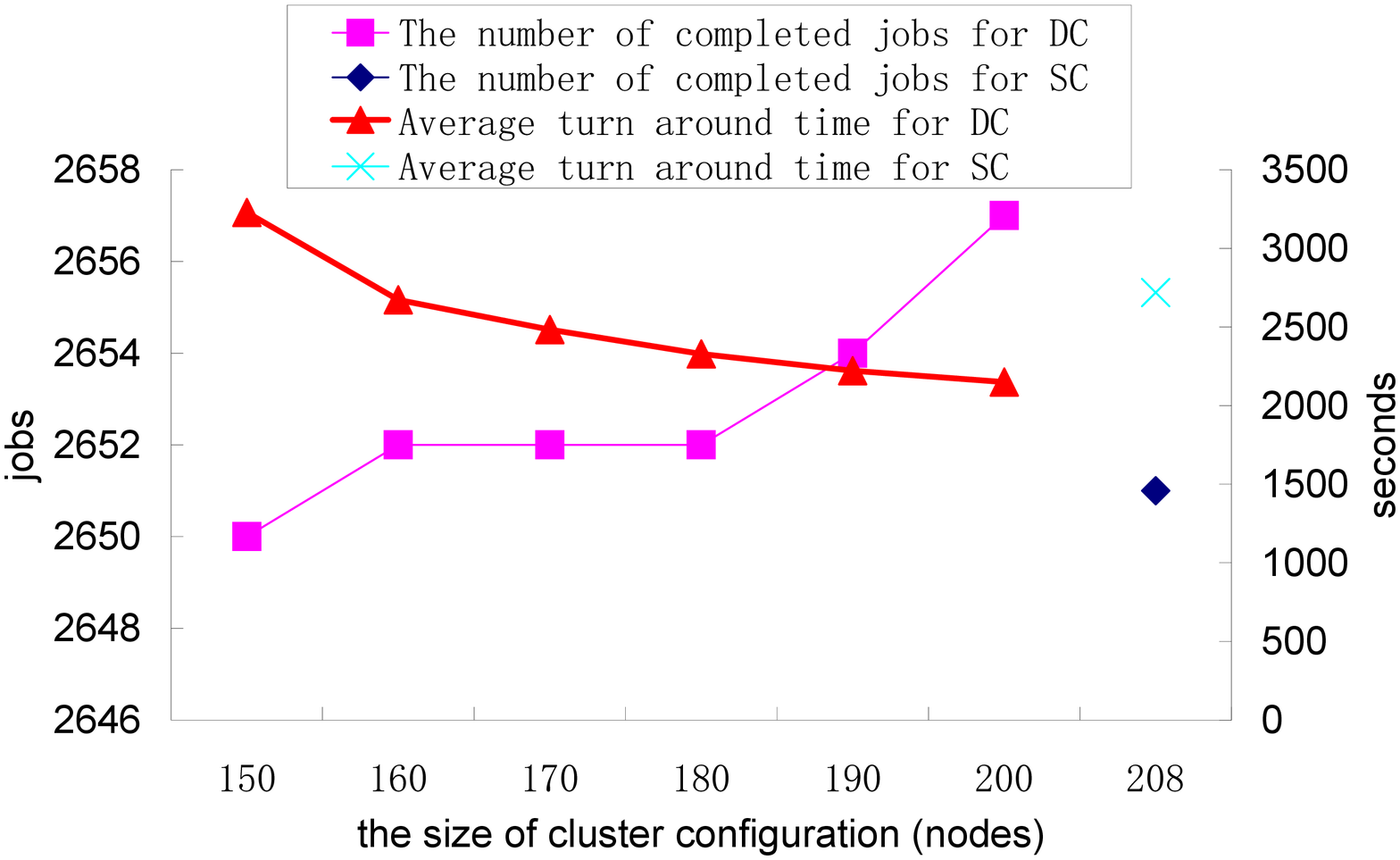}
\caption{For HPC trace, the number of completed jobs and the average turnaround time per jobs in two
weeks with different size of the cluster configuration.}
\label{first_experiment}
\end{figure}

In our simulation system, for \emph{static configuration} (in short \emph{SC}), of which each department of a large organization maintains its own cluster system, the minimum scale of the cluster system for HPC trace introduced in Section \ref{trace} is 144 nodes, because the real SDSC trace is also collected from the same 144 nodes; the minimum scale of the cluster system for Web service is 64 nodes, because the peak resource demand in Fig. \ref{varying_resource} is 64 virtual machines. So \emph{the size of the cluster configuration} allocated to Web services and HPC jobs for \emph{SC} is 208. For \emph{the dynamic configuration} (in short \emph{DC}), we respectively set \emph{the size of the cluster configuration} allocated to Web services and HPC jobs as 200, 190, 180, 170, 160 and 150. Fig. \ref{first_experiment} shows \emph{the number of completed jobs} and \emph{the average turnaround time per job in term of seconds} for HPC trace in two weeks when we set different \emph{size of the cluster configuration}.

For HPC trace, 2672 jobs are submitted to \emph{ST Server}. For \emph{dynamic configuration}, when the cost of the large organization in term of \emph{the size of the cluster configuration} decreases to 160, only 76.9\% of that of \emph{static configuration}, the benefit of scientific computing department in term of \emph{the number of completed jobs} in two weeks is still higher than that of \emph{static configuration}; while the benefit of end user in term of \emph{the reciprocal of the average turnaround time per job} is still higher than that of \emph{static configuration}.
With \emph{the size of the cluster configuration} decreases, \emph{the number of killed jobs} increases in general. Only the exception is the number of killed jobs when \emph{the size of the cluster configuration} is 170, which is higher than that when \emph{the size of the cluster configuration} is 160.

\begin{figure}[h]
\centering
\includegraphics[width=3.5in]{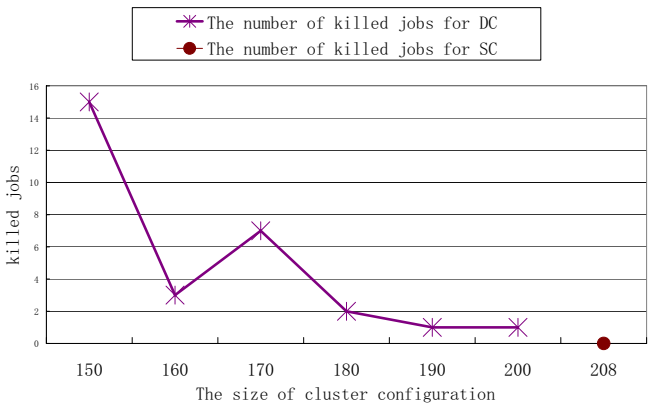}
\caption{For HPC trace, the number of killed jobs in two weeks with different size of the cluster configuration. }
\label{second_experiment}
\end{figure}
For Web service, the benefits of service providers and end users are unchanging, since we just use the same
resource consumption collected from Section \ref{resource_consumption} as the input to \emph{Resource Simulator}.

\section{ CONCLUSIONS} \label{conclusion}
Different departments of large organizations often maintain dedicated cluster systems for different computing loads. In this paper, we have designed and implemented a cloud management system software \emph{Phoenix Cloud} to consolidate HPC jobs and Web service application on the shared cluster system. We have also proposed cooperative resource provisioning and management policies of large organizations and their affiliated departments to share the consolidated cluster systems.
Our experiments show that in comparison with the case that each department of the same large organization runs its dedicated cluster system, consolidating HPC jobs and Web service applications from different departments with \emph{cooperative resource provisioning and management policies} not only significantly decreases the scale of the required cluster system for a large organization, but also improves the benefit of the scientific computing departments while provisioning enough resources to the other department running Web service applications with varying loads.

\section{ ACKNOWLEDGEMENTS} \label{acknowlegements}
This paper is supported by the National Science Foundation for Young Scientists of China (Grant
No. 60703020).



\end{document}